\documentstyle[preprint,aps,psfig]{revtex}
\begin{document}
\draft
\title{\bf Probing Black-Hole Physics in the Laboratory Using \\ High Intensity
Femtosecond Lasers}
\author{G. Sch\"afer and R. Sauerbrey}
\address{Theoretisch-Physikalisches Institut und
Institut f\"ur Optik und Quantenelektronik \\
Friedrich-Schiller-Universit\"at Jena, Max-Wien-Platz 1,
D-07743 Jena, Germany}
\date{\today}
\maketitle
\begin{abstract}
It is shown how laboratory experiments performed with
high intensity femtosecond lasers can probe the physics
of black holes in the near-horizon regime.
The acceleration generated by the high intensity laser ranging from $10^{13}$g
to more than $10^{18}$g
is identified with the gravitational acceleration at stretched horizons.
In the black-hole's asymptotic region, the stretched-horizon-reflected light
shows a measurable universal phase acceleration of $c^4/4GM$.

\end{abstract}

\bigskip

\bigskip

\bigskip

One of the major unresolved problems in astrophysics is the unique
identification of black holes. Due to the limited spatial resolution of
astronomical observations a large mass concentration alone is not sufficient.
Processes showing the existence of an event horizon are, however,
clear signatures of a black hole. In this letter we describe a scenario
where the investigation of light emission from stretched horizons of black
holes can be simulated in intense femtosecond laser-plasma interaction
experiments and simultaneously leads to a clear signature for a black hole.

Physics near the event horizon of black holes is known to be extreme
\cite{TPM86}.
The nearer to the event horizon matter and radiation are located the
stronger the gravitational attraction becomes that pulls them
to the horizon. In the event-horizon limit the accelerations grow
unboundedly. This contrasts sharply with extended bodies which at their
surfaces, according to general relativity,
can reach gravitational accelerations of about $10^{12}$g only \cite{W72},
where g denotes the acceleration at the surface of the
earth (g $\approx 10$m/s$^2$). Recently,
laboratory experiments employing table-top high intensity femtosecond
lasers have produced large accelerations over a wide range exceeding
$10^{18}$g,
and this even for macroscopic bodies with masses on the order of picograms
\cite{S96}, \cite{HTN97}. In this letter we shall consider highly
accelerated laboratory plasmas in correspondence with plasmas which may exist
near the event horizon of a rotating black hole immersed in a homogeneous
magnetic field. This analogy leads to a novel
possibility for exploring the near-horizon regime of black holes
by means of a phase sensitive analysis of the emitted radiation,
and thus, to uniquely identify black holes.

In the following discussion it will be sufficient to consider the
{\em geometry} of uncharged and
non-rotating black holes only. An object at rest in the gravitational field
of a black hole experiences a proper acceleration as given,
in Schwarzschild coordinates, by $b = (GM/r^2)(1-r_s/r)^{-1/2}$,
where $G$, $M$, and $r_s$ denote
the Newtonian gravitational constant, the black-hole mass, and the
Schwarzschild radius (radius of event horizon),
respectively, \cite{TPM86}. $r_s$ is given by $r_s=2GM/c^2$, where
$c$ denotes the speed of light. If the coordinate distance $\epsilon$
from the event horizon is much smaller than $r_s$, i.e. $0 < \epsilon
\lesssim 0.1 r_s$, the acceleration reads $b = c^2/2(r_s\epsilon)^{1/2}$.
Surfaces with radii $r=r_s+\epsilon$
are called {\em stretched} horizons. At stretched horizons
the gravitational acceleration is finite. In terms of the geometric length $l$
of the coordinate distance $\epsilon$, $l=2(r_s\epsilon)^{1/2}$, the acceleration
simply reads, $b=c^2/l$.

The maximum surface gravitational acceleration of an extended
object is reached for $\epsilon = r_s/8$ (Buchdahl limit).
This object might be a neutron star with roughly two solar masses.
Although no extended object can exist for radii smaller the $9r_s/8$
\cite{W72}, matter (plasma) can be kept at rest between the
Schwarzschild and Buchdahl radii of a black hole. Any mechanism that keeps the
plasma at rest at a stretched horizon will lead to large accelerations in the
plasma and consequently to a characteristic modification of the phase of an
electromagnetic wave emitted or reflected from this plasma.
The easiest way to keep plasma at rest in
this regime is to assume that the black hole is electricly charged. In an
astrophysical setting, however,
this charge may not to be large enough \cite{W84}. Much stronger
electric fields can be obtained by, e.g., a rotating black hole immersed in a
homogeneous magnetic field $B$, \cite{W74}.
Such an electric field corresponds to a charged black hole with net electric
charge equal to zero and with charge density at the event horizon's north and
south poles of $\sigma_{BJ} = -BJ/2\pi r_s^2c$. In order to achieve
sufficient charge densities a small angular momentum $J$, i.e.
$cJ/GM^2 \lesssim 0.1$, can be sufficient. The assumed geometry of an uncharged
and non-rotating hole is then still an excellent approximation.
Averaged over sufficiently long times, the charge density $\sigma_{BJ}$ is
neutralized by a charge density $\sigma_Q$ resulting from infalling charges,
i.e. the temporal average of the total charge density will vanish,
$\sigma_Q + \sigma_{BJ} = 0$. Because of stochastic
processes such as pair creation which keep the black hole's
magnetosphere filled with plasma \cite{TPM86}, and turbulences in the plasma,
we may, however, expect also large charge fluctuations $\delta \sigma_Q$
in the stretched horizon regime, with $|\delta \sigma_Q| << |\sigma_{BJ}|$.
For fluctuations corresponding to $M=2M_{\odot}$, $B=10^{10}$
gauss, and $cJ/GM^2 = 0.03$ ($M_{\odot}$ is the
solar mass) the resulting repulsive electric force cancels the
gravitational force acting on the incoming plasma ions, at a stretched horizon
located at $l=0.5$cm; i.e. $b=10^{18}$g. The temporal duration of the fluctuations must exceed
$l/c \approx 20$ ps to stop and halt the plasma long enough for the effect
to be measured. The magnetic field exceeding $10^{10}$ gauss, needed for $\sigma_{BJ}$,
may be supplied by an accreting plasma \cite{FKR92}. Of course, if $l$ is
increased, corresponding to larger stretched horizons, the charge fluctuation,
magnetic field or angular momentum of the black hole may be considerably
smaller to achieve the same effect.

It was shown recently that high intensity femtosecond lasers produce hot,
dense matter of solid density and keV-electron energies \cite{PM94}. Internal
pressures approaching Gigabars lead to a rapidly expanding plasma that
experiences accelerations on the order of $b \approx 10^{19}$ m/s$^2$
$\approx 10^{18}$ g. Similar accelerations are caused by the radiation pressure that also
reaches about 1 Gigabar at an intensity of
$I \approx 3 \times 10^{18}$ W/cm$^2$, \cite{HTN97}. These
accelerations actually dominate the motion of plasma material during the laser
pulse duration of $\tau_L \approx 100$ fs. For an expanding plasma
the final velocity reached by the plasma
corresponds to the ion acoustic velocity of $v_i \approx 10^7 ... 10^8$ cm/s
and is small compared to the speed of light. Due to the short laser-pulse
duration and the rapid heating of the plasma, however, this velocity is
reached in a very short time and thus the acceleration may be estimated to
$b \approx v_i/\tau_L \gtrsim 10^{18}$ m/s$^2$. Such accelerations have recently been
measured by spectral analysis of chirped laser pulses reflected from the
plasma surface \cite{S96}, \cite{HTN97}.

For a direct
measurement of the acceleration of the plasma surface (more precisely, the
location of the critical electron density), amplitude and phase of the
reflected laser pulse or a temporally delayed probe pulse have been experimentally
determined. Modifications in the reflected laser light of an ultrashort laser
pulse may be caused by kinematic effects, i.e. movement of the reflecting
surface as well as by self-phase modulation induced by rapidly changing
optical properties of the reflecting surface due to fast heating and
ionization of the plasma. We consider laser
pulses of pulse durations of $\tau_L \approx 100$ fs and intensities ranging
from $10^{16} ... 10^{20}$ W/cm$^2$. For typical aluminium plasma parameters ($Z
\approx 10$; $k_BT_e \approx 1$ keV; $M_i = 27$ amu) we obtain
for intensities in the $10^{16}$ W/cm$^2$ range and a high contrast ratio ($>10^9$)
of the laser pulse $L \lesssim v_i\tau_L \approx 20$ nm as an upper limit for
the plasma scale length. Therefore, $L/\lambda << 1$ holds for typical
laser wavelength of $\lambda \approx 800$ nm. The reflected wave
can therefore be obtained using the Fresnel approximation for steep density
gradients, and the kinematics of the plasma surface is described by a
``moving mirror'' \cite{LU92} of temporally varying reflectivity \cite{XX98}.

We approximate the mirror reflectivity by the following simple function for
the field amplitude, $R(t) = R_0 e^{-t/\tau_R}e^{i\phi}$, where $R_0$ is the
fraction of the reflected field amplitude, $\tau_R$ the characteristic time
for the decay of the reflectivity, and $\phi$ the phase angle. Measurements
and calculations show \cite{S96}, \cite{LSS93} that usually $\tau_R >> \tau_L$. Furthermore we put
$\phi \approx 0$ corresponding to reflection at the critical density. The
propagation pattern of the impinging field with frequency $\omega_0$
has the form $\mbox{exp} \{i(\omega_0/c)[ct + x]\}$.
The surface of the solid is located at $x = x_0$ before arrival of the laser
pulse and the location of the critical surface as a function of time
is given by $x(t) = x_0 + v_0 t + bt^2/2$, where $v_0$ is an arbitrary
initial expansion velocity and $b$ is the acceleration of the plasma front
(more exactly, that of the critical density).
The electric field reflected from the accelerated mirror, for $|v_0| << c$ and
$|b| t_r << c/2$, where $t_r \equiv t - (x-x_0)/c$ is the retarded time,
has now the following form \cite{FD76}:
$E_R(t_r) = R(t_r) E(t_r) \mbox{exp} \{i(\omega_0/c)[x_0 + (c + 2 v_0) t_r
+ b t_r^2]\}$, where $E(t_r)$ is the pulse envelope.
Note that the acceleration of the phase
is twice the mirror acceleration. In the Fresnel
approximation the wave vector $k$ of the reflected laser pulse is that of the
impinging one and independent of $x$ ($ck=\omega_0$).

It is evident from the above equations that light emitted from a uniformly
accelerated source shows linear chirp in frequency, i.e. a linear dependence
of frequency on time, when measured in the rest frame.
Consequently, the observation of chirp from ``moving mirrors'' can be used to
measure acceleration.
Chirp is measured using phase sensitive pulse diagnostics (FROG, Frequency
Resolved Optical Gating) for the ultrashort
laser pulse \cite{DTK94}. Such experiments are presently in progress. First
results \cite{HTN97} indicate plasma accelerations of $b\approx 2 \times 10^{19}$ m/s$^2$ $\approx
2 \times 10^{18}$g for a $10^{18}$ W/cm$^2$ Titanium Saphire laser of 100 fs pulse
duration impinging on a carbon surface. Fig. 1 shows the result of such an
experiment. From the FROG analysis the amplitude $E_{in}(t)$ and the phase
angle $\phi_{in}(t)$ of the incident laser pulse are obtained. A pulse width of
$\tau_L \approx 100$ fs and an almost constant phase $\phi_{in}(t)$ are evident for
the incident pulse. The reflected laser pulse clearly shows the predicted
parabolic behavior of the phase $\phi_{out}(t) \sim t^2$ and correspondingly
a temporally broadened field amplitude $E_{out}(t)$.
An indirect measurement of plasma acceleration of this magnitude was already
obtained from the spectral modification of chirped ultrashort
laser pulses reflected from solid surfaces \cite{S96}. In the following we demonstrate
that such an accelerated plasma mirror in Minkowski space shows a phase behavior
of light reflected from it that is identical to the phase behavior of light
reflected from a plasma sheet at rest in Schwarzschild spacetime in the
immediate neighbourhood of the black hole's event horizon (Fig. 2).

A linearly accelerated plasma mirror in flat spacetime (Fig. 2a) with
motion of the type $x(t) = (x_0^2 + c^2t^2)^{1/2}$, which
for $x_0 >> ct$ means $x(t) = x_0 + bt^2/2$ with acceleration $b = c^2/x_0$,
can be replaced by a plasma mirror at rest with
respect to a uniformly accelerated reference frame, i.e. as being at rest in
a so-called Rindler wedge \cite{BD82}, (Fig. 2b).
In our laboratory experiment,
the condition $x_0 >> ct$ always holds, i.e. the maximum mirror
velocities ($\dot x = c^2t/x_0$) are always much smaller then the speed of
light, indeed they are on the order of the ion acoustic velocity $v_i$
of about $10^{-3}c$ \cite{S96}.
In the limit, $x_0 << ct$, the quantum thermal radiation from the mirror
of temperature $T=bh/4\pi^2ck_B$ results, where $h$ denotes the Planck
constant, \cite{FD76}. This radiation is measured by an observer in inertial motion
and originates from an accelerated mirror which is approaching the velocity of
light.

An accelerated observer located in the Rindler wedge at $\xi_0$
measures a quantum thermal radiation (Unruh radiation) of
temperature $T$ as given above with $x_0 = e^{a\xi_0}/a$ (this equation
relates the space coordinates of the Rindler wedge and the non-accelerated
frame, at $t=0$), \cite{U76}. The Unruh radiation originates from the purely geometric
(causal) properties of the event horizon of the Rindler wedge and needs no
material basis (although matter falling through the horizon can be involved).
It corresponds to the thermal radiation of black holes \cite{DF77}.

Applying the Einstein equivalence principle, the high-acceleration
laboratory experiments can directly be transferred to stretched horizons
of black holes. The mirror, at rest in the
Rindler wedge, corresponds to a mirror located at a stretched horizon
and the laser source, at rest in the Minkowski space, corresponds to a radially
freely falling source near the stretched horizon, having zero velocity at the
time $t=0$. This scenario evolves as follows: In the equilibrium case,
$\sigma_Q + \sigma_{BJ} =0$, plasma is falling freely along the magnetic field axis.
On short times scales, charge fluctuations on deeper located stretched horizons
generate time-varying effective charge densities of fractions of
$\sigma_{BJ}$. Their high-frequency electric fields repel and support the
infalling (mirror) plasma. The overshooting and backfalling plasma serves
as a light source.

This situation is best described in radial Kruskal coordinates ($v$, $u$)
which in Schwarzschild spacetime play the r\^ole the quasi-cartesian coordinates
play in Minkowski space. The important relation between Kruskal and
radial Schwarzschild tortoise coordinates,
$r^* = r + r_s \ \mbox{ln} (r/r_s - 1)$, reads, $u^2 = v^2 +
\mbox{exp}(r^*/r_s)$, \cite{MTW73}.
At stretched horizons we get $u = u_0 (1 + v^2/u_0^2)^{1/2}$, where
$u_0 = (\epsilon e/r_s)^{1/2}$ is the position of the stretched horizon at
$t=v=0$ ($e=\mbox{exp}(1)$). For $v << u_0$, the phase of the electric field reflected from
the stretched horizon takes the form
$\mbox{exp}\{i\omega [u_0 + v_r + v_r^2/u_0]\}$, where $\omega$ is the
angular frequency in Kruskal time and $v_r \equiv v - u +
u_0$ is the retarded Kruskal time. The structure of the
phase as seen by an observer at rest in Kruskal coordinates
near the stretched horizon corresponds exactly to
the phase of the light reflected from the accelerated plasma mirror that
appears in $E_R(t)$, for $v_0=0$.
To know what an observer measures far away from the stretched horizon
we have to transform the phase of the wave to Schwarzschild coordinates.
If $r^*_0$ corresponds to $u_0$, i.e. $u_0 = \mbox{exp}(r^*_0/2r_s)$,
then $v_r = u_0 (1 - \mbox{exp}\{-ct_r/2r_s\})$ holds with $t_r \equiv t - (r^*
-r^*_0)/c$, and for the phase propagation follows,
taking into account $t_r << 2r_s/c$ $(\ge 40 \mu$s, for two solar masses as
minimum mass):
$\mbox{exp}(i\omega u_0) \mbox{exp}\{i(\omega u_0c/2r_s) [t_r +
ct_r^2/4r_s]\}$. In this expression, $\omega u_0c/2r_s$
and $b_F \equiv c^2/2r_s = c^4/4GM$ are respectively the angular frequency and the
phase acceleration measured with respect to the observer's proper time $t$
far away from the horizon. It is important to note that the measured
acceleration is independent from the chosen stretched horizon, i.e. it is the
same for all $\epsilon \lesssim 0.1 r_s$, resp. $b \gtrsim 10^{13}$g.
It is also worthwhile to mention that direct light from radially freely
backfalling sources in the stretched horizon regime, i.e. light not being
reflected at a stretched horizon, reveals an acceleration pattern in its
phase with just the opposite sign for the acceleration.

For a neutron star with mass $M$ the gravitational acceleration
at its surface is smaller than $(8/9)^2b_F$ so, using phase measurements,
a discrimination between black holes and neutron stars is possible if the mass
of the object is known with sufficient precision by other means. In the
neutron star case the reflecting mirror is made by its surface and the source
for light are near-surface plasma processes.

The expression for the acceleration of the phase occurs also in the
Unruh (purely outgoing) and Hawking radiation of a black hole with Hawking
temperature of $T=b_Fh/4\pi^2ck_B$ far away from the black hole's event horizon.
At the stretched horizons the Hawking temperature takes the
blueshifted value $T_{SH}=bh/4\pi^2ck_B$ with $b$ as given above.
The Unruh radiation originates from regions very close to the black-hole
event horizon after very long times ($t \rightarrow \infty$). The radiation
considered in this letter on the contrary originates from stretched horizons
and, thus, appears much earlier in the black-hole formation process.

In summary we find that the application of the equivalence principle and the
concept of plasma loaded stretched horizons to laboratory experiments
measuring large laser induced accelerations lead to a new experimental method
to identify black holes.

For critically reading the manuscript the authors thank S.~M.~Kopeikin.
The research was supported by the Max-Planck-Gesellschaft (Grant
02160-361-TG74) and by the Training and Mobility of Researchers program
(ERB 4061 PL 95-0765) of the European Community within the Superintense Laser
pulse-Solid Interaction network.

\newpage

{\bf FIGURE CAPTIONS}

Fig. 1.

A $10^{18}$ W/cm$^2$, $100$ fs Titanium Saphire laser is focussed
onto a fresh carbon target. The incident (dotted curves) and the reflected
(drawn curves) light are analyzed
with respect to the temporal behavior of amplitude (drawn lines) and phase
(dashed lines). The incident
light shows a pulsewidth of $\approx 100$ fs and a constant phase. The
reflected light is broadened due to the large phase ($> 6 \pi$) that is
accumulated because of the strong acceleration of the plasma mirror by the
large radiation pressure ($\approx 300$ Mbars) of the incident laser pulse.

Fig. 2.

Reference frames demonstrating the equivalence of an
accelerated plasma mirror with plasma sheet at rest in the neighbourhood of
the event horizon of a black hole. (a) Reflected light from an accelerated mirror as
observed by an observer at rest in Minkowski space. Equidistant temporal
intervals such as the period of an electromagnetic wave are transformed into
time intervals with decreasing duration for increasing time. This corresponds
to the linear chirp in frequency observed from a uniformly accelerated mirror (see Fig. 1).
(b) In the Rindler wedge ($ \eta$ time coordinate, $\xi$ space coordinate)
the mirror is at rest from the time $t= \eta =0$ on. Not shown is the past event horizon
($\eta = - \infty, \xi = - \infty$). (c) The Schwarzschild spacetime is depicted
in Kruskal coordinates ($v$ time coordinate, $u$ radial space coordinate).
The radially freely falling source of light rays is at rest in Kruskal
coordinates over a short period of time. The light rays reflected at the plasma
loaded stretched horizon are received at infinity (on graphical reasons,
the source is switched off before the reflected light rays do cross it).
Not shown are light rays which are directly emitted to infinity.
\newpage
\psfig{figure=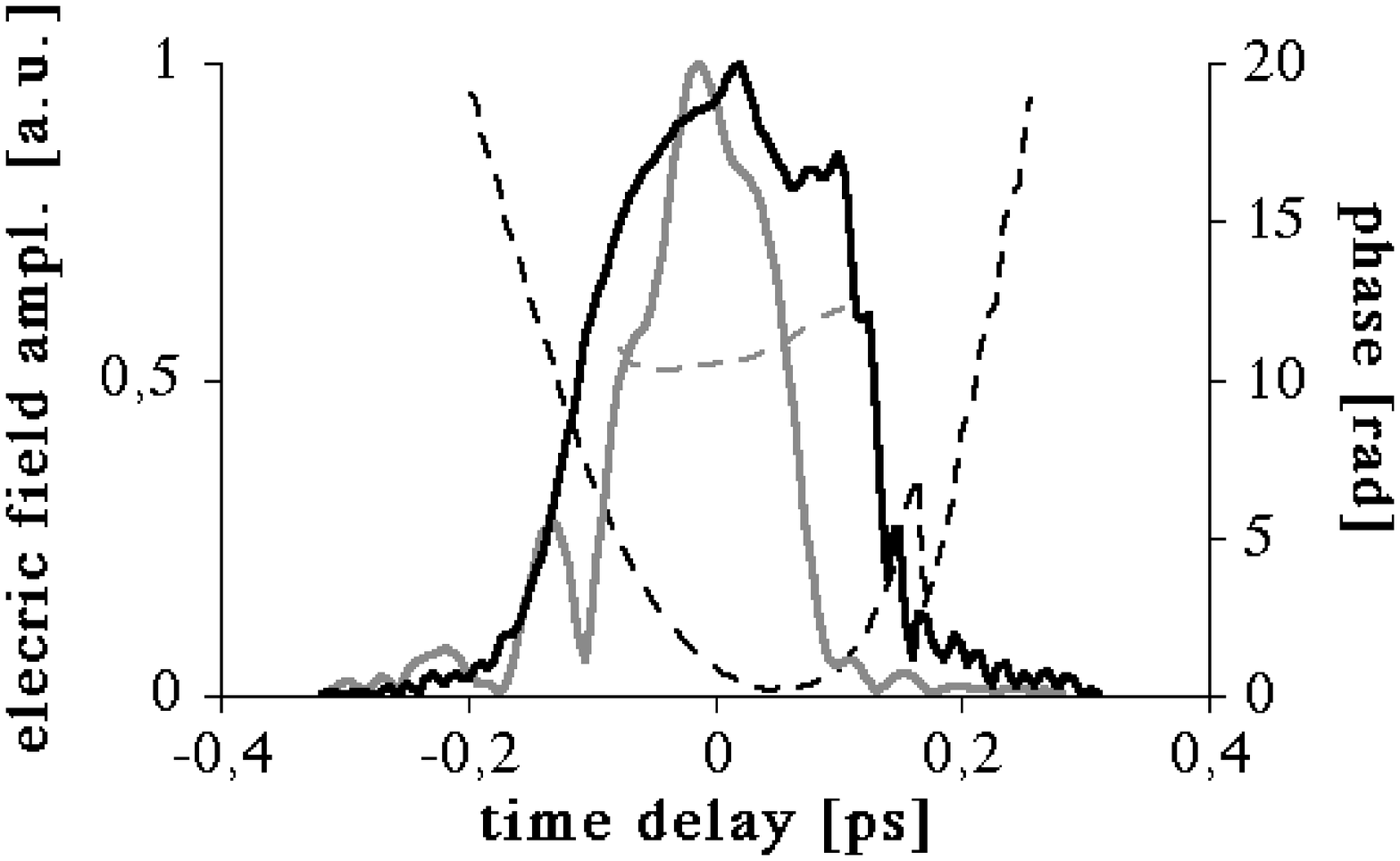,width=\textwidth}\newpage
\psfig{figure=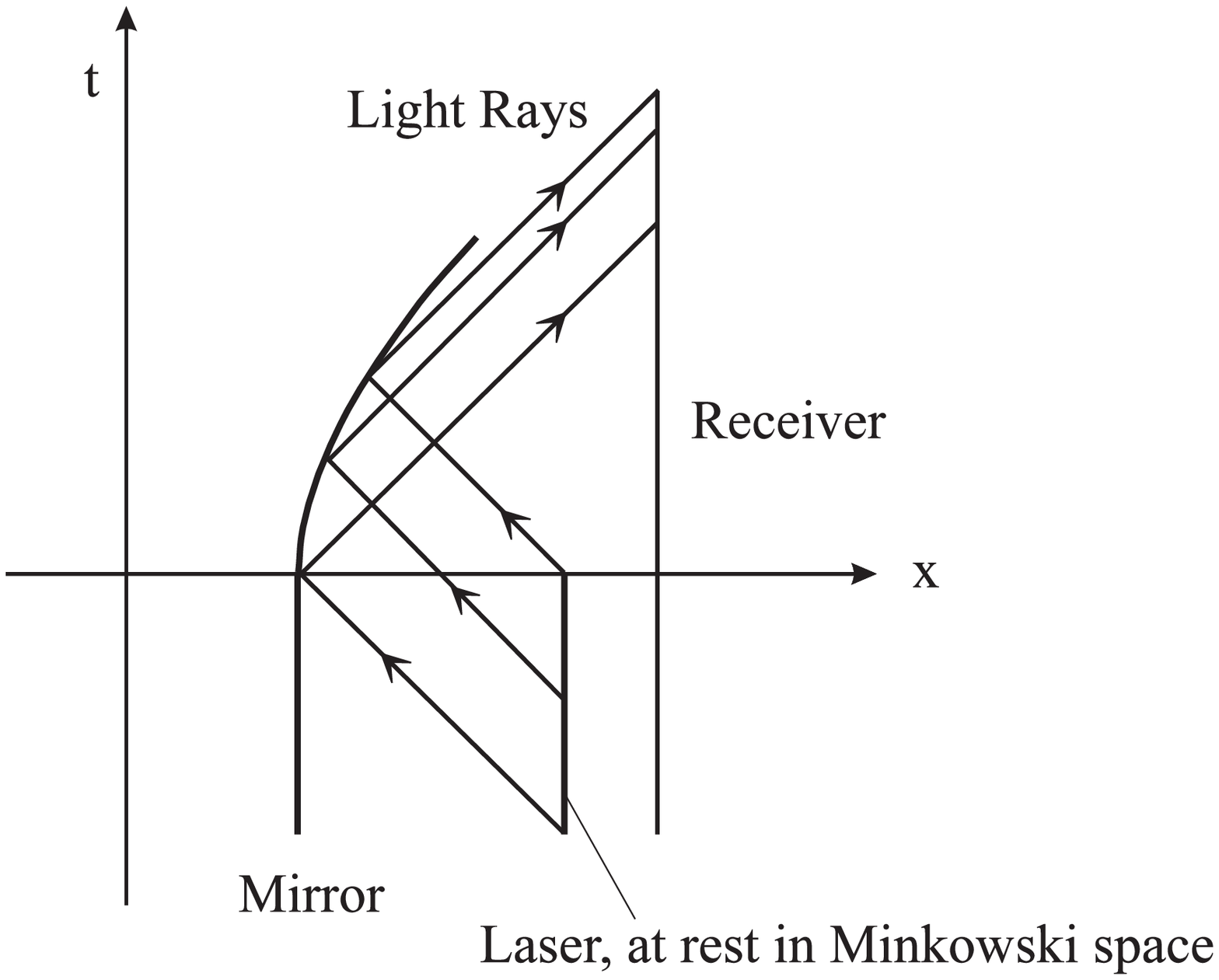,height=9.0cm,width=14.0cm}\vspace{2cm}
\psfig{figure=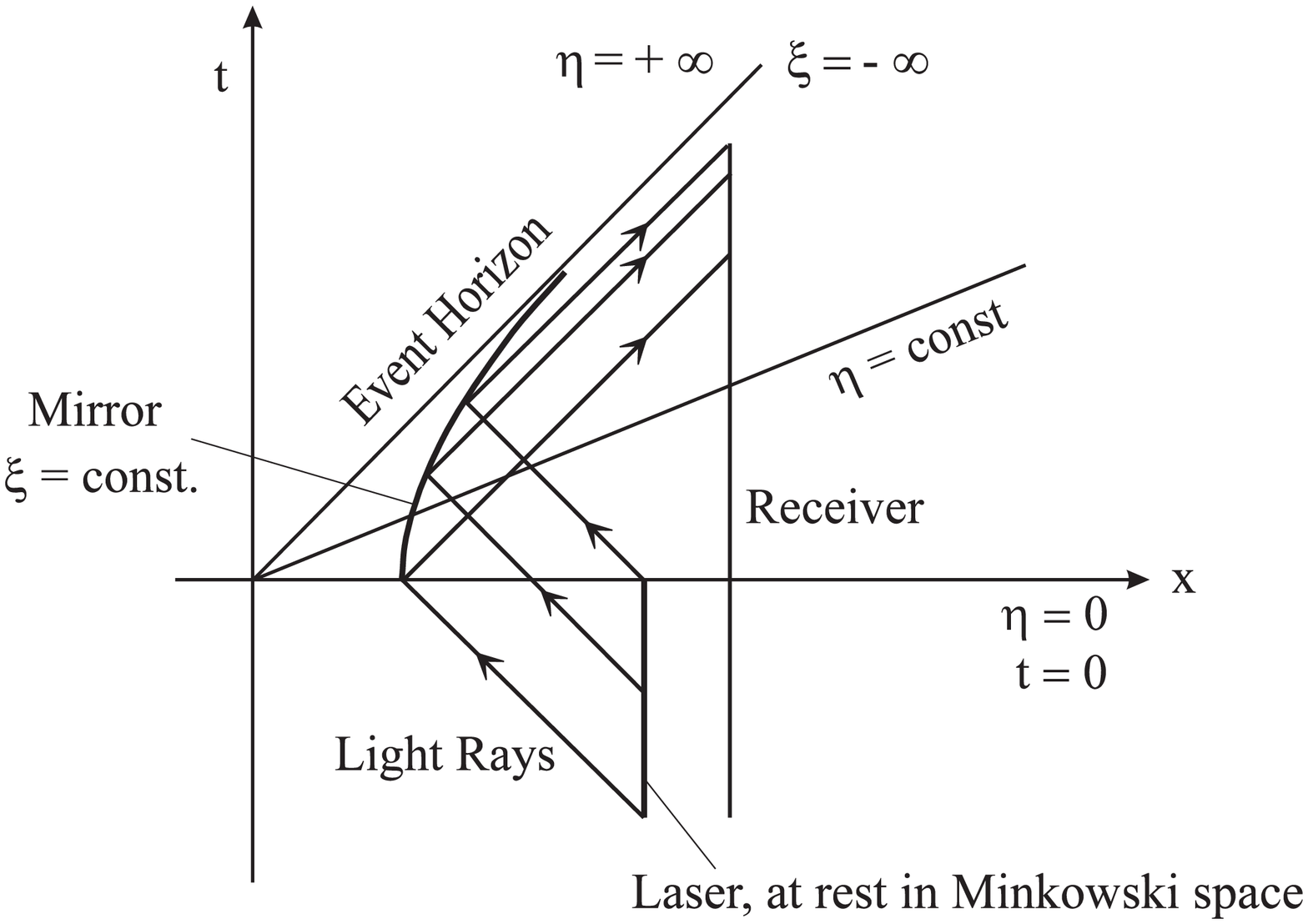,height=9.0cm,width=14.0cm}
\psfig{figure=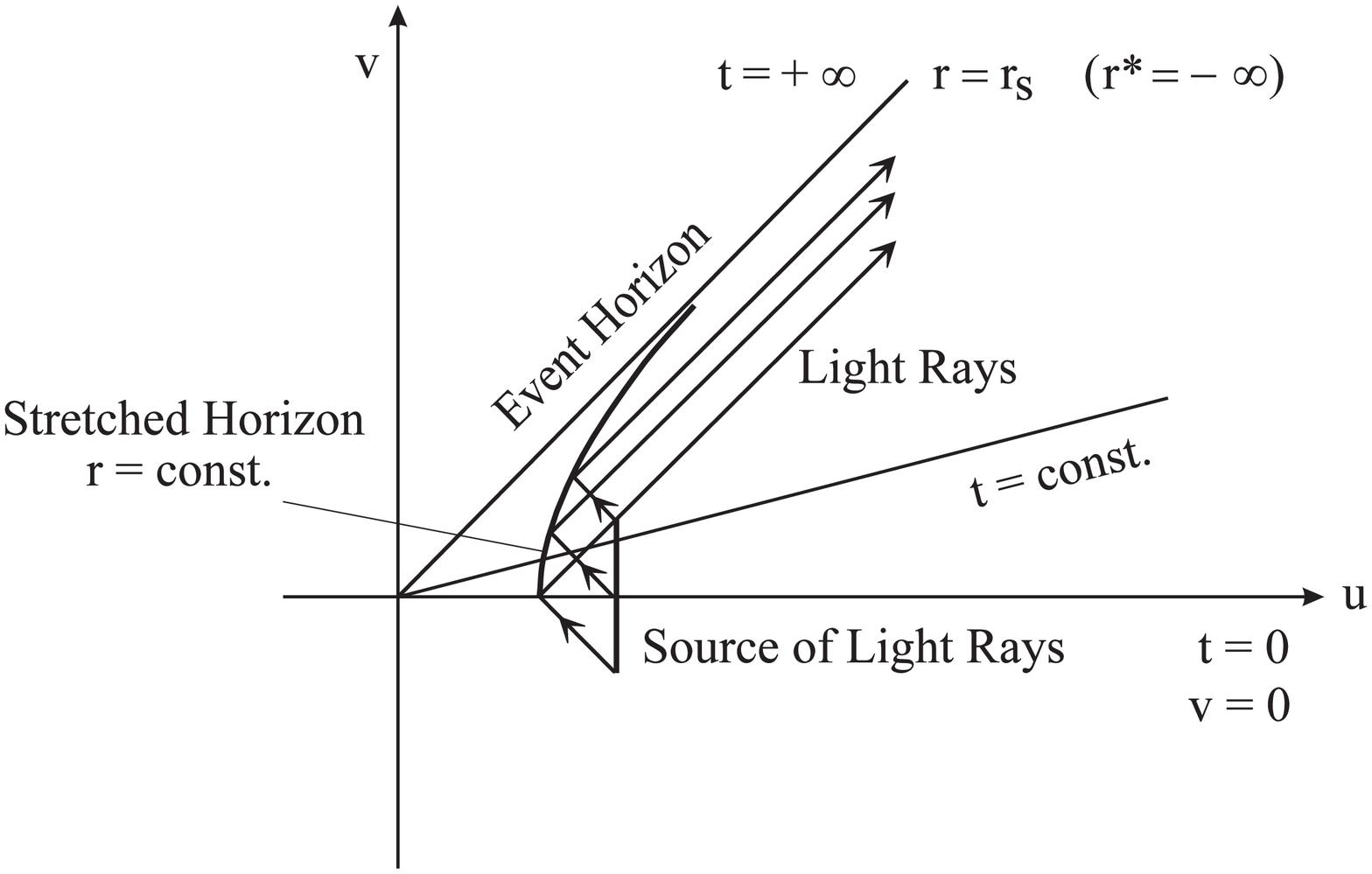,width=\textwidth}

\bigskip


\vspace{8mm}
\begin{references}


\bibitem{TPM86} K.~S.~Thorne, R.~H.~Price, and D.~A.~Macdonald (Eds.), {\em Black Holes: The
Membrane Paradigm} (Yale University Press, New Haven, 1986).

\bibitem{W72} S.~Weinberg, {\em Gravitation and Cosmology} (Wiley, New York,
1972), p. 334.

\bibitem{S96} R.~Sauerbrey, Phys. Plasmas {\bf 3}, 4712 (1996). In  Eq. (3)
of this paper the term $kx(t)$ must be multiplied by a factor of 2, which
transforms all values for velocities and accelerations to half the values
given in the paper without affecting the basic physical conclusions.

\bibitem{HTN97} R.~H\"a{\ss}ner, W.~Theobald, S.~Niedermeier, K.~Michelmann,
H.~Schillinger, T.~Feurer, G.~Sch\"afer, and R.~Sauerbrey, in {\em
Proceedings of the International Conference on Superstrong Fields in Plasmas}
(Varenna, Italy, August 27 - September 2, 1997).

\bibitem{W84} R.~M.~Wald, {\em General Relativity} (The University of Chicago
Press, Chicago, 1984), p.~314.

\bibitem{W74} R.~M.~Wald, {\em Phys.~Rev.~D} {\bf 10}, 1680 (1974).

\bibitem{FKR92} J.~Frank, A.~King, and D.~Raine, {\em Accretion Power in
Astrophysics} (Cambridge University Press, Cambridge, 1992), p.~238.

\bibitem{PM94} M.~D.~Perry and G.~Mourou, Science {\bf 264}, 917 (1994).


\bibitem{LU92} X.~Liu and D.~Umstadter, Phys.~Rev.~Lett. {\bf 69}, 1935 (1992).


\bibitem{XX98}
Since the electron density and consequently the plasma frequency
first increases during the impinging laser pulse and later decreases due to
recombination, the reflectivity of the plasma surface is a rapidly varying
function of time. It can be shown that for perpendicular incidence when the
light is reflected from the vicinity critical surface the phase shift due to the
varying electron density vanishes. Only the magnitude of the reflectivity
changes. This change in reflectivity has been measured using pump-probe
techniques \cite{LSS93}. It was shown that the reflectivity changes rapidly during the
plasma generating pulse and decays on a timescale $\tau_R$ of several
picoseconds.

\bibitem{LSS93} D.~v.~d.~Linde, H.~Schuler, H.~Schulz, and T.~Engers,
in {\em Ultrafast Phenomena VIII}, Springer
Series in Chemical Physics (Springer-Verlag, Berlin, 1993), Vol.~55, p.~280.


\bibitem{FD76} S.~A.~Fulling and P.~C.~W.~Davies, {\em Proc.~R.~Soc.~London}
{\bf A348}, 393 (1976).

\bibitem{BD82} N.~D.~Birrell and P.~C.~W.~Davies, {\em Quantum Fields in Curved Space}
(Cambridge University Press, Cambridge, 1982).

\bibitem{U76} W.~G.~Unruh, {\em Phys.~Rev.~D} {\bf 14}, 870 (1976).

\bibitem{DF77} P.~C.~W.~Davies and S.~A.~Fulling, {\em Proc.~R.~Soc.~London}
{\bf A356}, 237 (1977).
\bibitem{DTK94} K.~W.~DeLong, R.~Trebino, and D.~J.~Kane, {\em J.~Opt.~Soc.~Am.}
B {\bf 11}, 1595 (1994).

\bibitem{MTW73} C.~W.~Misner, K.~S.~Thorne, and J.~A.~Wheeler, {\em Gravitation}
(Freeman, San Francisco 1973), pp. 827 - 835.



\end{references}
\end{document}